%% file: ms.tex
\documentclass[conference]{IEEEtran}

\input{sec-preamble.tex}

\usepackage{enumitem}

\definecolor{uclablue}{RGB}{39,116,174}
\definecolor{uclabluedarkest}{RGB}{0,59,92}
\definecolor{uclabluedarker}{RGB}{0,85,135}
\definecolor{uclabluelighter}{RGB}{139,184,232}
\definecolor{uclabluelightest}{RGB}{218,235,254}

\definecolor{uclagold}{RGB}{255,209,0}
\definecolor{uclagolddarker}{RGB}{255,199,44}
\definecolor{uclagolddarkest}{RGB}{255,184,28}

\definecolor{uclaviolet}{RGB}{130,55,165}
\definecolor{uclamagenta}{RGB}{255,0,165}

\usepackage{pgfplots}
\pgfplotsset{compat=newest}
\usetikzlibrary{plotmarks}
\usetikzlibrary{arrows.meta}
\usepgfplotslibrary{patchplots}
\usepackage{grffile}

\pgfplotsset{plot coordinates/math parser=false}

\begin{document}

\input{sec-title.tex}

\input{sec-abstract.tex}

\glsresetall

\input{sec-introduction.tex}

\input{sec-system-model.tex}

\input{sec-contribution.tex}

\input{sec-numerical-results.tex}

\input{sec-conclusion.tex}

\input{sec-acknowledgment}

\input{sec-bibliography.tex}

\end{document}

%% file: sec-preamble.tex
\usepackage{cite}

  \usepackage[pdftex]{graphicx}

\usepackage{amsmath}
\usepackage{algorithm}
\usepackage{algorithmic}

  \usepackage[caption=false,font=footnotesize]{subfig}
\usepackage{url}

\usepackage{amsthm}

\newtheoremstyle{mydefinition}
{}
{}
{}
{0pt}
{\bfseries}
{.}
{ }
{\thmname{#1}\thmnumber{ #2}: \thmnote{#3}}

\theoremstyle{mydefinition}

\newtheoremstyle{myremark}
{}
{}
{}
{0pt}
{\bfseries}
{.}
{ }
{\thmname{#1}\thmnumber{ #2}: \thmnote{#3}}
\theoremstyle{myremark}

\newtheoremstyle{remarkshort}
{}
{}
{}
{0pt}
{\bfseries}
{.}
{ }
{\thmname{#1}\thmnumber{ #2}}

\theoremstyle{remarkshort}

\theoremstyle{remarkshort}

\input{math-ipr.tex}

\input{glossary.tex}

\newcommand{\figref}[1]{Fig.~\ref{#1}}

\usepackage{mathtools}

\usepackage[top=0.751in, bottom=1.04in, left=0.625in, right=0.625in]{geometry}


%% file: math-ipr.tex
\newcommand{\comment}[1]{{}}

\usepackage{amssymb}
\usepackage{xspace}
\usepackage{bm}
\usepackage{bbm}

\let\originalleft\left
\let\originalright\right
\renewcommand{\left}{\mathopen{}\mathclose\bgroup\originalleft}
\renewcommand{\right}{\aftergroup\egroup\originalright}

\newcommand{\mat}[1]{\ensuremath{\mathbf{#1}}\xspace} 
\renewcommand{\vec}[1]{\ensuremath{\mathbf{#1}}\xspace} 

\newcommand{\real}{\ensuremath{\mathbb{R}}\xspace}

\newcommand{\setreal}{\ensuremath{\real}}

\def\vb{{\vec{b}}}
\def\vc{{\vec{c}}}

\def\ve{{\vec{e}}}

\def\vg{{\vec{g}}}
\def\vh{{\vec{h}}}

\def\vp{{\vec{p}}}

\def\vw{{\vec{w}}}

\def\vz{{\vec{z}}}

\def\mA{{\mat{A}}}

\def\mW{{\mat{W}}}



%% file: glossary.tex
\usepackage{xspace}
\usepackage[acronym,nogroupskip,nonumberlist,nopostdot]{glossaries}
\makeglossaries
\usepackage{relsize}

\usepackage{xcolor}

\newacronym{snr}{SNR}{signal-to-noise ratio}
\newacronym{sinr}{SINR}{signal-to-interference-plus-noise ratio}
\newacronym{inr}{INR}{interference-to-noise ratio}
\newacronym{sir}{SIR}{signal-to-interference ratio}
\newacronym{sqr}{SQR}{signal-to-quantization-noise ratio}
\newacronym{sqnr}{SQNR}{signal-to-quantization-plus-noise ratio}
\newacronym{ian}{IAN}{interference as noise}
\newacronym{ber}{BER}{bit error rate}
\newacronym{pn}{PN}{pseudorandom noise}
\newacronym{bfsk}{BFSK}{binary frequency shift keying}
\newacronym{fh}{FH}{frequency-hopped}
\newacronym{fh-bfsk}{FH-BFSK}{frequency-hopped binary frequency shift keying}
\newacronym{crc}{CRC}{cyclic redundancy check}
\newacronym{isi}{ISI}{intersymbol interference}
\newacronym{dsss}{DSSS}{direct-sequence spread spectrum}
\newacronym{ofdm}{OFDM}{orthogonal frequency-division multiplexing}
\newacronym{ofdma}{OFDMA}{orthogonal frequency-division multiple access}
\newacronym{sdr}{SDR}{software-defined radio}
\newacronym{tx}{TX}{transmitter}
\newacronym{rx}{RX}{receiver}
\newacronym{fdd}{FDD}{frequency-division duplexing}
\newacronym{tdd}{TDD}{time-division duplexing}
\newacronym{fdma}{FDMA}{frequency-division multiple access}
\newacronym{tdma}{TDMA}{time-division multiple access}
\newacronym{sdma}{SDMA}{space-division multiple access}
\newacronym[plural=MPCs]{mpc}{MPC}{multipath component}
\newacronym{mui}{MUI}{multi-user interference}
\newacronym{lsb}{LSB}{least significant bit}
\newacronym{jcas}{JCAS}{joint communication and sensing}

\newacronym{qam}{QAM}{quadrature amplitude modulation}
\newacronym{mqam}{MQAM}{M-ary quadrature amplitude modulation}

\newacronym{ls}{LS}{least-squares}
\newacronym{lms}{LMS}{least mean squares}
\newacronym{rls}{RLS}{recursive least-squares}
\newacronym{rzf}{RZF}{regularized zero-forcing}
\newacronym{mmse}{MMSE}{minimum mean square error}
\newacronym{lmmse}{LMMSE}{linear minimum mean square error}
\newacronym{mse}{MSE}{mean square error}
\newacronym{fft}{FFT}{fast Fourier transform}
\newacronym{dft}{DFT}{discrete Fourier transform}
\newacronym{dtft}{DTFT}{discrete-time Fourier transform}
\newacronym{ctft}{CTFT}{continuous-time Fourier transform}
\newacronym{ml}{ML}{machine learning}
\newacronym[plural=NNs]{nn}{NN}{neural network}
\newacronym[plural=RNNs]{rnn}{RNN}{recurrent neural network}
\newacronym[plural=ADCs]{adc}{ADC}{analog-to-digital converter}
\newacronym[plural=DACs]{dac}{DAC}{digital-to-analog converter}
\newacronym[plural=FPGAs]{fpga}{FPGA}{field-programmable gate array}
\newacronym{evm}{EVM}{error vector magnitude}
\newacronym{enob}{ENOB}{effective number of bits}
\newacronym{zf}{ZF}{zero-forcing}
\newacronym{rv}{r.v.}{random variable}
\newacronym{omp}{OMP}{orthogonal matching pursuit}
\newacronym{svd}{SVD}{singular value decomposition}

\newacronym{sdp}{SDP}{semidefinite programming}
\newacronym{psd}{PSD}{positive semidefinite}
\newacronym{nsd}{NSD}{negative semidefinite}

\newacronym{ks}{K-S}{Kolmogorov-Smirnov}

\newacronym{mad}{MAD}{median absolute deviation around the median}

\newacronym{agc}{AGC}{automatic gain control}
\newacronym{rf}{RF}{radio frequency}
\newacronym{if}{IF}{intermediate frequency}
\newacronym{los}{LOS}{line-of-sight}
\newacronym{nlos}{NLOS}{non-line-of-sight}
\newacronym{ple}{PLE}{path loss exponent}
\newacronym[plural=dB,firstplural=decibels (dB)]{db}{dB}{decibel}
\newacronym[plural=dBm,firstplural=decibel milliwatts (dBm)]{dbm}{dBm}{decibel milliwatts}
\newacronym{pa}{PA}{power amplifier}
\newacronym{lna}{LNA}{low noise amplifier}
\newacronym{vga}{VGA}{variable gain amplifier}
\newacronym{cw}{CW}{continuous wave}
\newacronym{papr}{PAPR}{peak-to-average power ratio}
\newacronym{usrp}{USRP}{Universal Software Radio Peripheral}
\newacronym{irr}{IRR}{image rejection ratio}
\newacronym{lo}{LO}{local oscillator}
\newacronym{vm}{VM}{vector modulator}
\newacronym{mmwave}{mmWave}{millimeter wave}
\newacronym{eirp}{EIRP}{effective isotropic radiated power}
\newacronym{rsrp}{RSRP}{reference signal received power}

\newacronym{csma}{CSMA}{carrier-sense multiple access}
\newacronym{csmaca}{CSMA/CA}{carrier-sense multiple access with collision avoidance}
\newacronym{csmacd}{CSMA/CD}{carrier-sense multiple access with collision detection}
\newacronym{mac}{MAC}{medium access control}
\newacronym{phy}{PHY}{physical layer}
\newacronym{4g}{4G}{fourth generation}
\newacronym{lte}{LTE}{Long-Term Evolution}
\newacronym{4glte}{4G LTE}{\gls{4g} \gls{lte}}
\newacronym{5g}{5G}{fifth generation}
\newacronym{nr}{NR}{New Radio}
\newacronym{5gnr}{5G NR}{5G New Radio}
\newacronym{ieee}{IEEE}{Institute of Electrical and Electronics Engineers}
\newacronym{wifi}{Wi-Fi}{IEEE 802.11}
\newacronym{lan}{LAN}{local area network}
\newacronym{wlan}{WLAN}{wireless local area network}
\newacronym[plural=BSs]{bs}{BS}{base station}
\newacronym[plural=SBSs]{sbs}{SBS}{small-cell base station}
\newacronym[plural=FD-SBSs]{fdsbs}{FD-SBS}{\gls{fd}-enabled \gls{sbs}}
\newacronym[plural=MBSs]{mbs}{MBS}{macrocell base station}
\newacronym[plural=UEs]{ue}{UE}{user equipment}
\newacronym{ul}{UL}{uplink}
\newacronym{dl}{DL}{downlink}
\newacronym{qos}{QoS}{Quality of Service}
\newacronym{fcc}{FCC}{Federal Communications Commission}
\newacronym{iab}{IAB}{integrated access and backhaul}
\newacronym{fab}{FAB}{fixed access and backhaul}
\newacronym{hetnet}{HetNet}{heterogeneous network}

\newacronym{siso}{SISO}{single-input single-output}
\newacronym{mimo}{MIMO}{multiple input multiple output}
\newacronym{sumimo}{SU-MIMO}{single-user \gls{mimo}}
\newacronym{mumimo}{MU-MIMO}{multi-user \gls{mimo}}
\newacronym{bf}{BF}{beamforming}
\newacronym{ca}{CA}{constant amplitude}
\newacronym{ula}{ULA}{uniform linear array}
\newacronym{upa}{UPA}{uniform planar array}
\newacronym[\glslongpluralkey={angles of arrival}]{aoa}{AoA}{angle of arrival}
\newacronym[\glslongpluralkey={angles of departure}]{aod}{AoD}{angle of departure}
\newacronym{dof}{DoF}{degrees of freedom}
\newacronym{csi}{CSI}{channel state information}
\newacronym{csit}{CSIT}{\gls{csi} at the transmitter}
\newacronym{csir}{CSIR}{\gls{csi} at the receiver}
\newacronym{cs}{CS}{compressed sensing}

\newacronym{fd}{FD}{in-band full-duplex}
\newacronym{hd}{HD}{half-duplex}
\newacronym{si}{SI}{self-interference}
\newacronym{sic}{SIC}{self-interference cancellation}
\newacronym{soi}{SoI}{signal of interest}
\newacronym{asic}{A-SIC}{analog \acrlong{sic}}
\newacronym{dsic}{D-SIC}{digital \gls{sic}}
\newacronym{star}{STAR}{simultaneous transmit and receive}
\newacronym{warp}{WARP}{Wireless Open-Access Research Platform}
\newacronym{bfc}{BFC}{beamforming cancellation}
\newacronym{ipi}{IPI}{inter-panel-interference}
\newacronym{ipic}{IPIC}{inter-panel-interference cancellation}

\newacronym{qcqp}{QCQP}{quadratically-constrained quadratic programming}
\newacronym{pdf}{PDF}{probability density function}
\newacronym{cdf}{CDF}{cumulative density function}
\newacronym{iid}{i.i.d.}{independently and identically distributed}

\newacronym{elf}{ELF}{extremely low frequency}
\newacronym{slf}{SLF}{super low frequency}
\newacronym{ulf}{ULF}{ultra low frequency}
\newacronym{vlf}{VLF}{very low frequency}
\newacronym{lf}{LF}{low frequency}
\newacronym{mf}{MF}{medium frequency}
\newacronym{hf}{HF}{high frequency}
\newacronym{vhf}{VHF}{very high frequency}
\newacronym{uhf}{UHF}{ultra high frequency}
\newacronym{shf}{SHF}{super high frequency}
\newacronym{ehf}{EHF}{extremely high frequency}
\newacronym{thf}{THF}{tremendously high frequency}

\newacronym{leo}{LEO}{low-earth orbit}

\newacronym{wncg}{WNCG}{Wireless Networking and Communications Group}
\newacronym{linc}{LINC}{Laboratory of Informatics, Networks, and Communications}
\newacronym{ut}{UT Austin}{The University of Texas at Austin}
\newacronym{uiuc}{UIUC}{University of Illinois at Urbana-Champaign}
\newacronym{usc}{USC}{University of Southern California}
\newacronym{mit}{MIT}{Massachusetts Institute of Technology}
\newacronym{berkeley}{UC Berkeley}{University of California, Berkeley}
\newacronym{osu}{OSU}{Ohio State University}



%% file: sec-title.tex
\title{Satellite Assignment Policy Learning for Coexistence in LEO Networks}

\author{
    \IEEEauthorblockN{Jeong Min Kong\textsuperscript{1}, Eunsun Kim\textsuperscript{2}, Ian P.~Roberts\textsuperscript{1}}
    \IEEEauthorblockA{\textsuperscript{1}Wireless Lab, Department of Electrical and Computer Engineering, UCLA, Los Angeles, CA, USA}
    \IEEEauthorblockA{\textsuperscript{2}6G@UT, Wireless Networking and Communications Group, University of Texas at Austin, Austin, TX, USA\\
    Email: \{jeongminkong, ianroberts\}@ucla.edu}
}

\maketitle

%% file: sec-abstract.tex
\begin{abstract}
Unlike in terrestrial cellular networks, certain frequency bands for low-earth orbit (LEO) satellite systems have thus far been allocated on a non-exclusive basis. In this context, systems that launch their satellites earlier---referred to as \emph{primary} systems---are given spectrum access priority over those that launch later, known as \emph{secondary} systems. For a secondary system to function, it is expected to either coordinate with primary systems or ensure that it does not cause excessive interference to primary ground users. Reliably meeting this interference constraint requires real-time knowledge of the receive beams of primary users, which in turn depends on the primary satellite-to-primary user associations. However, in practice, primary systems have thus far not publicly disclosed their satellite assignment policies; therefore, it becomes essential for secondary systems to develop methods to infer such policies. Assuming there is limited historical data indicating which primary satellites have served which primary users, we propose an end-to-end graph structure learning-based algorithm for learning highest elevation primary satellite assignment policies, that, upon deployment, can directly map the primary satellite coordinates into assignment decisions for the primary users. Simulation results show that our method can outperform the best baseline, achieving approximately a 15\% improvement in prediction accuracy.
\end{abstract}


%% file: sec-introduction.tex
\section{Introduction} \label{sec:introduction}

Dense LEO satellite constellations are becoming key enablers of global broadband wireless connectivity \cite{leo_gen2,sat_6g,leo_cell,daesik_sat}. With Starlink already launching over 6,000 satellites to date \cite{sat_stats}, and other companies, such as Amazon Kuiper, also recently sending their first set of satellites to orbit \cite{kuiper_first_launch}, it is becoming evident that LEO networks will play an even more crucial role in providing universal broadband access in the future.

Unlike traditional terrestrial cellular networks, where spectrum is typically allocated exclusively, the Federal Communications Commission (FCC) and other spectrum authorities have assigned certain frequency bands to LEO constellations in a non-exclusive manner \cite{47cfr25_261,itur1323}. To ensure fair and efficient sharing of this spectrum, the FCC prioritizes systems that submitted launch applications in earlier so-called processing rounds, designated as \emph{primary} systems; as a result, it is expected that newer systems, referred to as \emph{secondary} systems, operate by either coordinating with or protecting primary systems~\cite{fcc_nrules}. In this context, an interference protection constraint is often defined as restricting the secondary system's interference so that it increases the effective noise temperature at the primary ground user's receiver by no more than 6\%, which corresponds to an interference-to-noise ratio (INR) of no greater than $-12.2$~dB \cite{ntia_ipc,itur1323}.

The work of \cite{feasibility_journal} has recently shown that, regardless of how the primary system serves its ground users, a secondary system can, in principle, operate without causing any interference beyond the protection constraint, as there will always be at least one secondary satellite capable of avoiding conflict with any primary user. Building on this finding, \cite{coexistence_arxiv} introduced a framework for selecting the optimal secondary satellites that can maximize the downlink capacity of the secondary users while guaranteeing protection of the primary users. One of the key assumptions underlying their approach is knowledge of the primary satellite assignments---that is, knowing which primary satellite is serving which primary user at any given time. This information is critical, as understanding the receive beam directions of the primary users is crucial for identifying secondary satellites that can avoid causing excessive interference. In practice, however, primary satellite association policies---such as that of Starlink---are proprietary and remain unknown to secondary operators like Amazon Kuiper and Omnispace. Given the absence of regulatory requirements mandating the disclosure of such policies \cite{coexistence_arxiv}, it becomes increasingly important to develop practical methods for determining primary satellite association policies without any direct input from the primary systems.

A novel deep learning (DL)-based scheme for learning primary satellite assignment policies was presented in \cite{coexistence_arxiv}. Their approach involves first leveraging received signal power measurements from the primary satellites, collected by secondary users, to deduce which primary satellites are serving specific regions over time. This inferred data is then used to train a DL model whose objective is to predict the primary satellite serving any given location in the future, using only the publicly available positions of the primary satellites. While the authors proposed several simple DL methods for learning the highest elevation (HE) and maximum contact time (MCT) primary satellite assignment policies, they all resulted in limited prediction accuracies. Nevertheless, their findings showed that even a modest improvement in accuracy could substantially improve the performance of secondary satellite selection. Motivated by this insight, in this paper, we introduce a more advanced graph structure learning (GSL)-based approach for learning HE primary satellite assignment policies. Simulations show that our method can outperform existing baseline models in \cite{coexistence_arxiv}, delivering around 15\% improvement in prediction accuracy against their best baseline.

%% file: sec-system-model.tex
\section{System Model} \label{sec:system-model}

In this paper, we consider a scenario where satellites across two LEO constellations seek to use the same frequency band for downlink transmission to their respective ground users, distributed in an overlapping region.

In both systems, each satellite simultaneously serves its ground users through highly focused spot beams. Each beam illuminates a specific area on the Earth's surface, known as a \emph{cell}, which---like those in terrestrial cellular networks---is considered to have a hexagonal shape \cite{starlink_map,kuiper}. Because a satellite can form several beams at once, it is assumed to be responsible for serving a \emph{cluster} of $N_{\mathrm{C}}$ cells during the interval between successive handovers, which tends to last on the order of 15 seconds \cite{starlink_measure}. While the composition of each cell cluster remains fixed over time, the satellite assigned to serve a particular cluster will change according to the corresponding satellite association policy. In this work, we assume that the primary system follows a plausible, HE protocol for the satellite assignment.

Let there be $N_{\mathrm{G}}$ cell clusters in the region of interest, and $N_{\mathrm{P}}^t$ number of available primary satellites at time $t$. Among all of the primary satellites in orbit, a satellite is considered ``available'' if the elevation angle relative to the horizon of some designated reference point in the region is greater than a specified threshold \cite{47cfr25_261}; at all times, the number of available satellites will be between $1$ and $N_{\mathrm{A}}$. Under the assumed HE satellite assignment protocol, each cell cluster is assigned a unique priority level between $1$ and $N_{\mathrm{G}}$, with $1$ indicating the highest priority and $N_{\mathrm{G}}$ indicating the lowest priority. Let us further define $\mathcal{G} = \{\vg_{n}: n = 1,\dots,N_{\mathrm{G}}\}$ as a set representing all clusters, with $\vg_{n}$ denoting the $n$-th highest priority cluster of $N_{\mathrm{C}}$ ground cells $\vg_{n} = \{\vc_{n,\ell}: \ell = 1,\dots,N_{\mathrm{C}}\}$, and $\mathcal{\bar{P}}^t = \{\vp_{i}^t: i = 1,\dots, N_{\mathrm{P}}^t\}$ as a set representing all primary satellites at time $t$, with $\vp_{i}^t$ denoting the primary satellite with the $i$-th highest elevation angle relative to the designated reference point at time $t$.

Under the HE protocol, at every handover time, each cell cluster initially identifies all of the satellites whose minimum elevation angle, with respect to any point within the cluster, surpasses a predefined threshold. Formally, let $\mathcal{P}_{n}^t = \{\vp \in \mathcal{\bar{P}}^t: \epsilon(\vg_{n},\vp) \geq \epsilon_{\min}\}$ denote all satellites in $\mathcal{\bar{P}}^t$ whose minimum elevation angle $\epsilon$, measured from any location within the cluster $\vg_{n}$, exceeds some threshold $\epsilon_{\min}$. With these sets determined at all clusters, the satellite assignment then proceeds sequentially based on cluster priority, i.e., $\vg_{1}, \vg_{2}, \dots, \vg_{N_{\mathrm{G}}}$. At cluster $\vg_{k}$, $k = 1,\dots,N_{\mathrm{G}}$, the satellite in $\mathcal{P}_{k}^t$ with the highest elevation angle relative to its cluster center is first selected. If this satellite has already been assigned to some higher-priority cluster $\vg_{m}, 1 \leq m < k$, the next-highest elevation satellite in $\mathcal{P}_{k}^t$ is considered; this process continues until an unassigned satellite is found. If all of the satellites in $\mathcal{P}_{k}^t$ are occupied, then the one with the highest elevation angle is assigned by default.

%% file: sec-contribution.tex
\section{Graph Structure Learning-Based \\ Satellite Assignment Policy Learning} \label{sec:contribution}

\begin{figure}[!t]
    \centering
    \vspace{0.0in}    \includegraphics[width=\linewidth,height=\textheight,keepaspectratio]{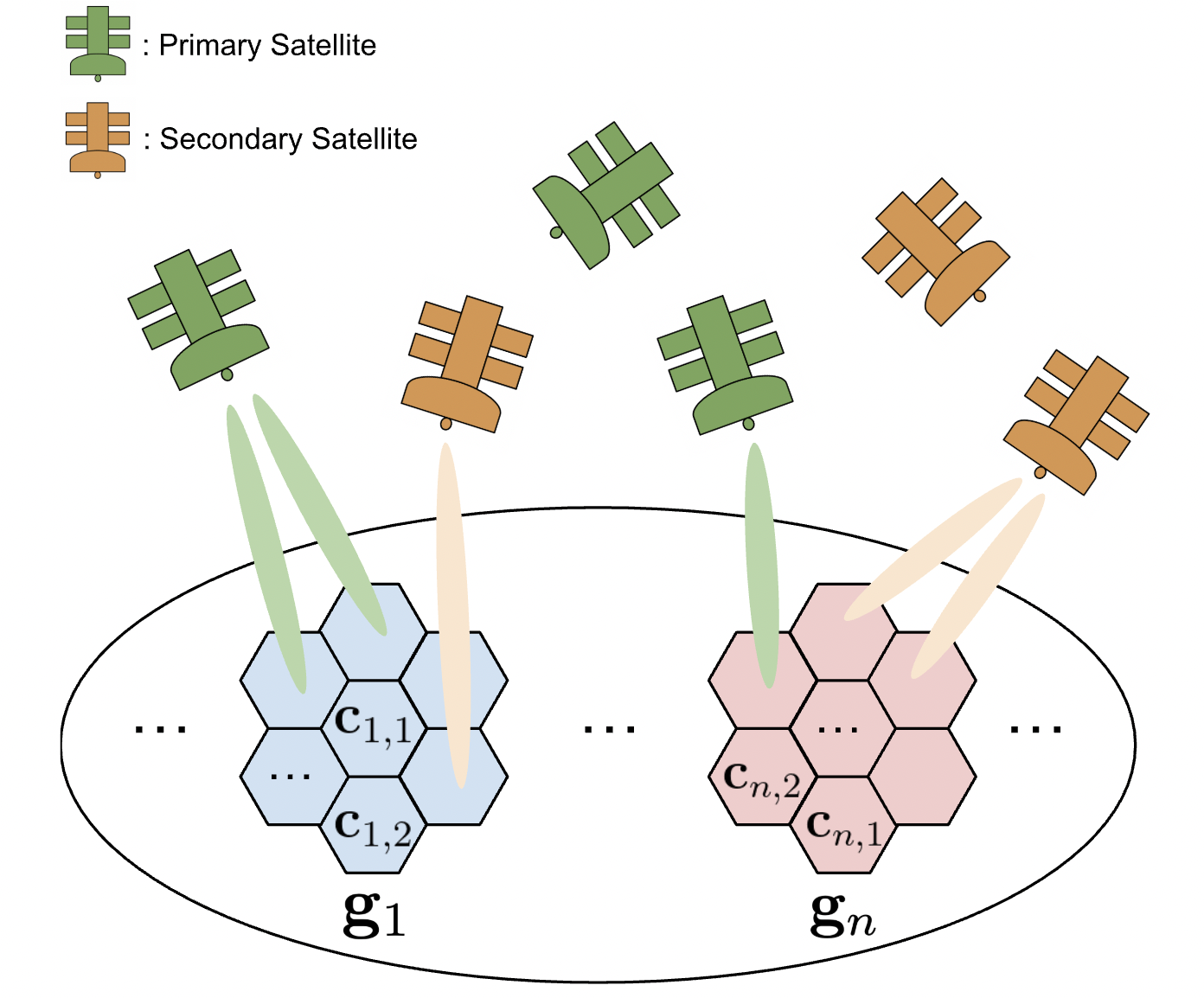}
    \caption{Primary and secondary satellites serving their respective ground users in the same frequency band using highly focused spot beams.}
    \label{fig:Figure System Model}
\end{figure}

\begin{figure*}[!t]
    \centering
    \includegraphics[width=0.98\linewidth]{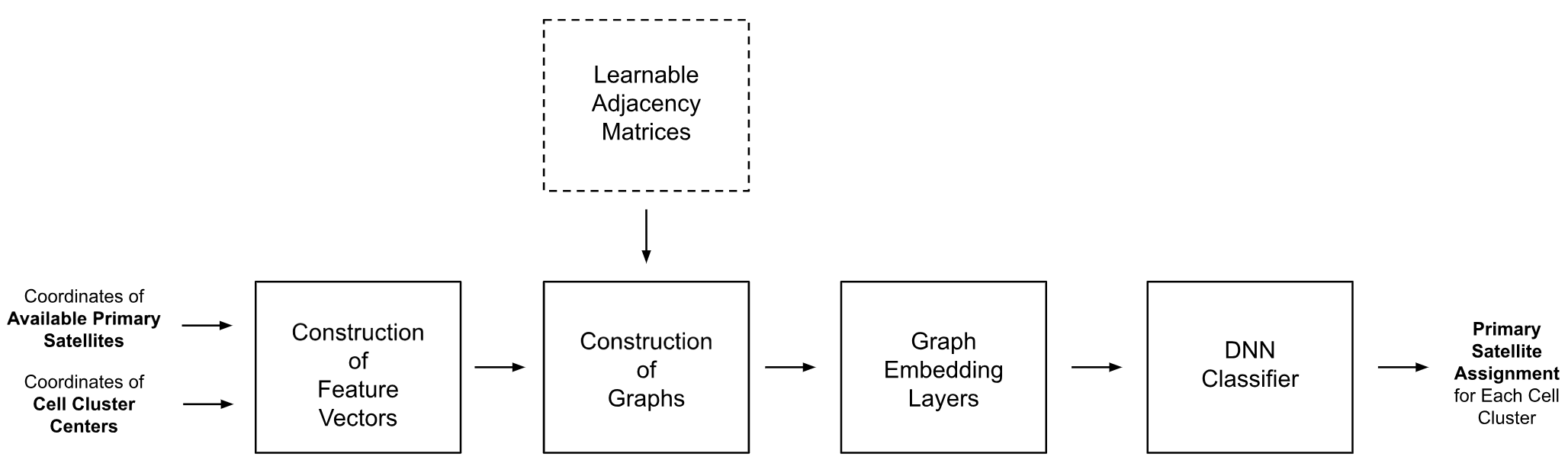}
    \caption{Block diagram of our proposed GSL-based primary satellite assignment policy learning approach.}
    \label{fig:Figure Method Overview}
\end{figure*}

As discussed, in real-world settings, secondary systems are unaware of the satellite assignment policy used by the primary system. In particular, when the primary system employs the HE satellite assignment protocol, the secondary systems lack knowledge of the \emph{priorities} assigned to the cell clusters. Assuming there is access to data indicating which primary satellites have been serving which primary users over some period of time---derived from the received signal powers measured by the secondary users---we propose an end-to-end GSL-based algorithm that aims to learn the unknown priority structure of the clusters, and, upon deployment, can directly map publicly available primary satellite coordinates into assignment decision for the clusters. We begin with a brief overview of the specific GSL architecture that we utilize before detailing our proposed approach.

\subsection{Graph Structure Learning Module}

GSL extends traditional graph neural networks (GNNs) by learning both the graph structure and the node representations simultaneously, instead of relying on a predefined topology~\cite{gsl_arxiv}. Unlike traditional GNN models such as graph convolutional networks (GCNs) and graph attention networks (GATs), which depend on fixed edge connections for message passing, GSL is designed for settings where the graph structure may be incomplete, noisy, or unavailable. Below, we outline the GSL framework utilized in this study.

Formally, consider a directed graph $G = (\mathcal{V},\mA)$, where $\mathcal{V}$ is the set of nodes and $\mA \in \setreal^{|\mathcal{V}| \times |\mathcal{V}|}$ is a \emph{learnable} adjacency matrix that defines the connectivity of the graph. Each node $i \in \mathcal{V}$ is initially represented by a vector known as the input feature vector $\vh_{i} \in \setreal^{F}$, where $F$ is defined as the number of input features. When these are passed through what is called a graph embedding layer, it generates a new representation called the output feature vector $\vh_{i}^{\prime} \in \setreal^{F^{\prime}}$ for each node, with $F^{\prime}$ denoting the number of output features, that encodes contextual information from itself and other relevant nodes in the graph.

Specifically, in a graph embedding layer, the input feature vector $\vh_{i}$ of each node $i \in \mathcal{V}$ is first linearly transformed using a shared learnable matrix $\mW \in \setreal^{F^{\prime} \times F}$ to form
\begin{align}
    \vz_{i} = \mW \vh_{i} \in \setreal^{F^{\prime}}.
\end{align}
Let $\bar{a}_{i,j} \in \mA$ represent the raw adjacency value of the directed edge from node $j \in \mathcal{V}$ to node $i$. Since this value is unbounded, i.e., can take any real number, it is constrained between 0 and 1 by passing it through the sigmoid function; let $a_{i,j}$ denote this normalized value. The output representation of each node $i$ is the \emph{weighted aggregation} of all nodes' transformed features, including itself, i.e.,
\begin{align}
    \vh_{i}^{\prime} = \sum_{j \in \mathcal{V}}^{} a_{i,j} \vz_{j} \in \setreal^{F^{\prime}},
\end{align}
where $a_{i,j}$ can be interpreted as the degree of importance that node $i$ assigns to the features of node $j$. Note that a non-linearity is not applied here, unlike in many graph learning algorithms.

To enhance the model's capacity to capture diverse perspectives from node features with reduced noise, multiple adjacency matrices and graph embedding layers, each with its own set of learnable parameters, can be employed in parallel. Formally, let $G^{(k)} = (\mathcal{V},\mA^{(k)})$, $k = 1,\dots,K$, represent graphs with the same features but each with a unique learnable adjacency matrix $\mA^{(k)}$. For each corresponding embedding layer $k$, a unique weight matrix $\mW^{(k)}$ is used to compute the transformed node features $\vz_{i}^{(k)} = \mW^{(k)} \vh_{i}$ independently. The output of each embedding layer is an aggregated feature representation for each node $i$, based on the weighted sum of all nodes under that embedding layer:
\begin{align}
    \vh_{i}^{\prime \ (k)} = \sum_{j \in \mathcal{V}}^{} a_{i, j}^{(k)} \vz_{j}^{(k)} \in \setreal^{F^{\prime}},
\end{align}
where $a_{i, j}^{(k)}$ denotes the normalized adjacency value from node $j$ to node $i$ in $G^{(k)}$. The outputs from all embedding layers are then combined to form the final representations,
\begin{align}
    \vh_{i}^{\prime} = ||_{k = 1}^{K} \vh_{i}^{\prime \ (k)} \in \setreal^{KF^{\prime}},
\end{align}
where $||$ denotes concatenation.

\subsection{Overall Model Architecture}

Here, we present our end-to-end GSL-based algorithm, designed to learn the unknown priorities and directly translate primary satellite coordinates into assignment decisions. A block diagram of this approach is shown in \figref{fig:Figure Method Overview}.

Suppose a secondary system wants to know the primary satellite assignments at time $t$. To begin, for each cluster $\vg_{n} \in \mathcal{G}$, we compute a vector $\bar{\vw}_{n}^{t} = (\bar{w}_{n,1}^{t},\bar{w}_{n,2}^{t},\dots,\bar{w}_{n,N_{\mathrm{A}}}^{t})$, where $\bar{w}_{n,i}^{t}$ represents the elevation angle between the center position of $\vg_n$ and satellite $\vp_{i}^{t} \in \mathcal{\bar{P}}^t$ for $1 \leq i \leq N_{\mathrm{P}}^t$, and 0 otherwise. To better represent this angular information, we apply the well-known cos/sin encoding to $\bar{\vw}_{n}^{t}$ to form $\tilde{\vw}_{n}^{t} = (\tilde{w}_{n,1}^{t},\tilde{w}_{n,2}^{t},\dots,\tilde{w}_{n,2N_{\mathrm{A}}}^{t})$, where $\tilde{w}_{n,2i-1}^{t} = \cos(\bar{w}_{n,i}^{t})$ and $\tilde{w}_{n,2i}^{t} = \sin(\bar{w}_{n,i}^{t})$ for $1 \leq i \leq N_{\mathrm{P}}^t$, and 0 otherwise. A one-hot vector $\ve_{n} \in \setreal^{N_{\mathrm{G}}}$, with a $1$ at the $n$-th index and $0$ elsewhere, is further concatenated with $\tilde{\vw}_{n}^{t}$ to serve as a unique identifier for cluster $\vg_{n}$. To enhance the expressiveness of the resulting representation, the combined vector $\tilde{\vw}_{n}^{t} || \ve_{n}$ is passed through a shared learnable non-linear layer that doubles its input dimensionality, leading to the final representation 
\begin{align}
    \vw_{n}^{t} = \sigma (\bar{\mW}(\tilde{\vw}_{n}^{t} || \ve_{n}) + \bar{\vb}) \in \setreal^{2(2N_{\mathrm{A}}+N_{\mathrm{G}})},
\end{align}
where $\sigma (\cdot)$ is the ReLU activation function, and $\bar{\mW}$ and $\bar{\vb}$ are the weights and biases of the linear layer, respectively.

After, we construct $K$ graphs with $N_{\mathrm{G}}$ nodes. For all graphs, each node represents a unique cell cluster, and the node corresponding to cluster $\vg_{n}$ is assigned with the feature vector $\vw_{n}^{t}$. Each graph's connectivity is defined by its unique learnable adjacency matrix. Once the graph structures are established, they are first passed through the $K$ graph embedding layers; the updated feature representations are then fed into a shared deep neural network (DNN), which outputs a vector of size $N_{\mathrm{A}}$ for each node. Each element in this vector represents a value proportional to the probability of assignment to a specific satellite, and the satellite corresponding to the highest value is selected as the serving satellite for that cluster. During training, the non-linear layer, adjacency matrices, graph embedding layers, and DNN are all jointly optimized to minimize the average cross-entropy loss across all clusters.

In essence, the main objectives of this approach can be summarized as follows. For each cluster/node $\vg_{n}$,
\begin{enumerate}
    \item find all clusters that have a higher priority, i.e., $\{ \vg_{k}: k = 1,\dots,n-1 \}$, and
    \item assign the correct satellite after carefully considering the priorities and angular information of itself and higher-priority clusters identified in 1).
\end{enumerate}
The motivation behind objective 1) is that only the satellite assignments at higher-priority clusters can impact the decision at $\vg_n$. Information from lower-priority clusters is largely irrelevant, as higher-priority clusters will always take precedence in cases where satellite assignments overlap. In our algorithm, this step is implemented using learnable adjacency matrices, where the value of the edge from cluster $j$ to cluster $i$ can be interpreted, in part, as a confidence score that cluster $j$ holds a higher priority than cluster $i$ (for $j \neq i$).

To better illustrate objective 2), refer to \figref{fig:Figure Example}, where the goal is to determine the correct satellite assignment for the third-highest priority cluster, $\vg_{3}$. For simplicity, assume that each element in the feature vector represents the elevation angle between the considering cluster and the satellite corresponding to the element's index. In this example, satellite 2 is assigned to $\vg_{1}$ since the elevation angle corresponding to satellite 2 is the largest and $\vg_{1}$ has the highest priority. Ideally, $\vg_{2}$, which has the second highest priority, would also be assigned satellite 2 since it has the highest angle; however, because it is already taken by $\vg_{1}$, $\vg_{2}$ is instead assigned the next best option---satellite 3. Similarly, $\vg_{3}$ would ideally take satellite 3, then satellite 2, but since both are already occupied, it settles for satellite 4. To achieve such a correct selection, the algorithm needs more than just the angular information from higher-priority clusters---it must also accurately identify their priorities and use all of this knowledge to \emph{reason} and make the right decisions in a sequential, priority-aware manner. To enable this, we jointly leverage a set of learnable linear transformation matrices in the graph embedding layers---each acting as a unique encoder of angular and priority information---and a DNN classifier, that together form a unified architecture for determining the most likely satellite assignment. \\

%% file: sec-numerical-results.tex
\section{Evaluation} \label{sec:numerical-results}

\begin{figure}[!t]
    \centering
    \vspace{0.0in}%
    \includegraphics[width=\linewidth]{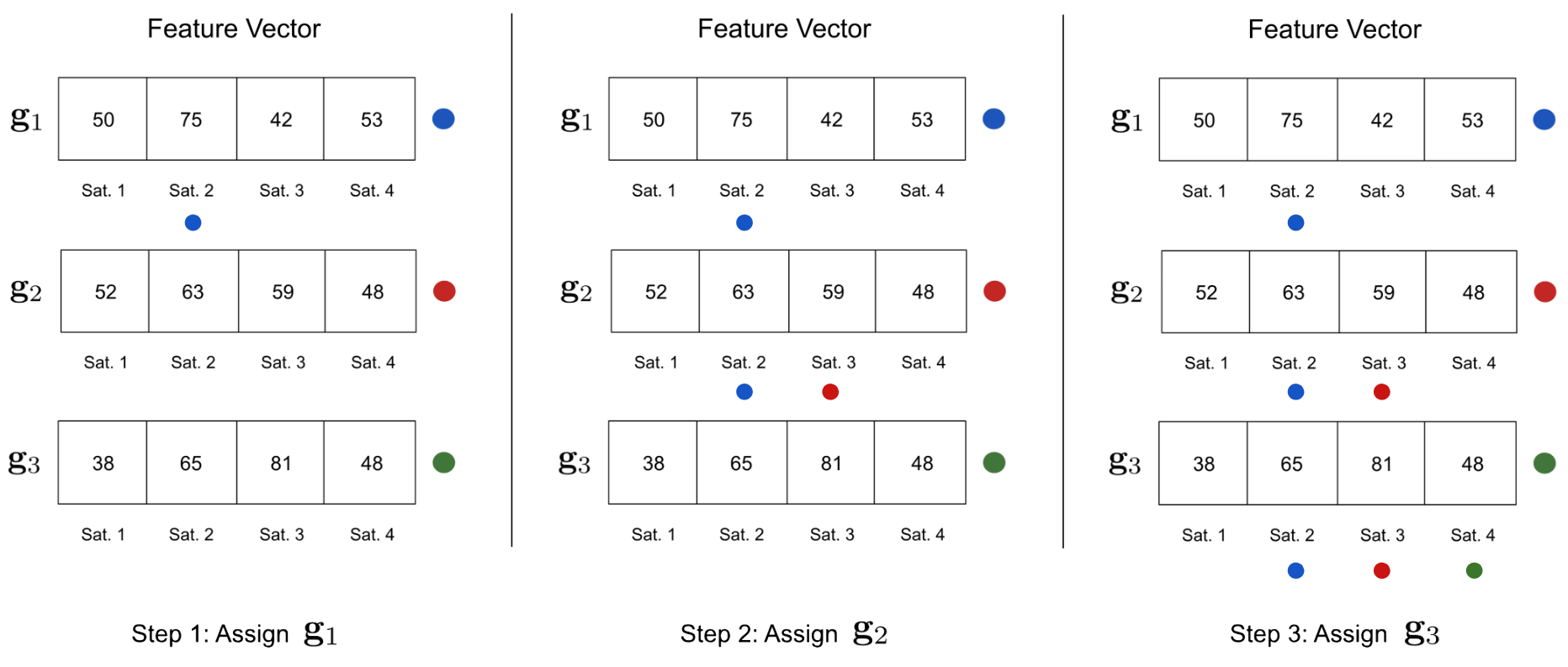}
    \caption{An example demonstrating satellite assignments under the HE protocol. It begins with cluster $\vg_{1}$, which holds the highest priority, followed sequentially by $\vg_{2}$ and then $\vg_{3}$.}
    \label{fig:Figure Example}
\end{figure}

\begin{figure}[!t]
    \centering
    \includegraphics[width=\linewidth]{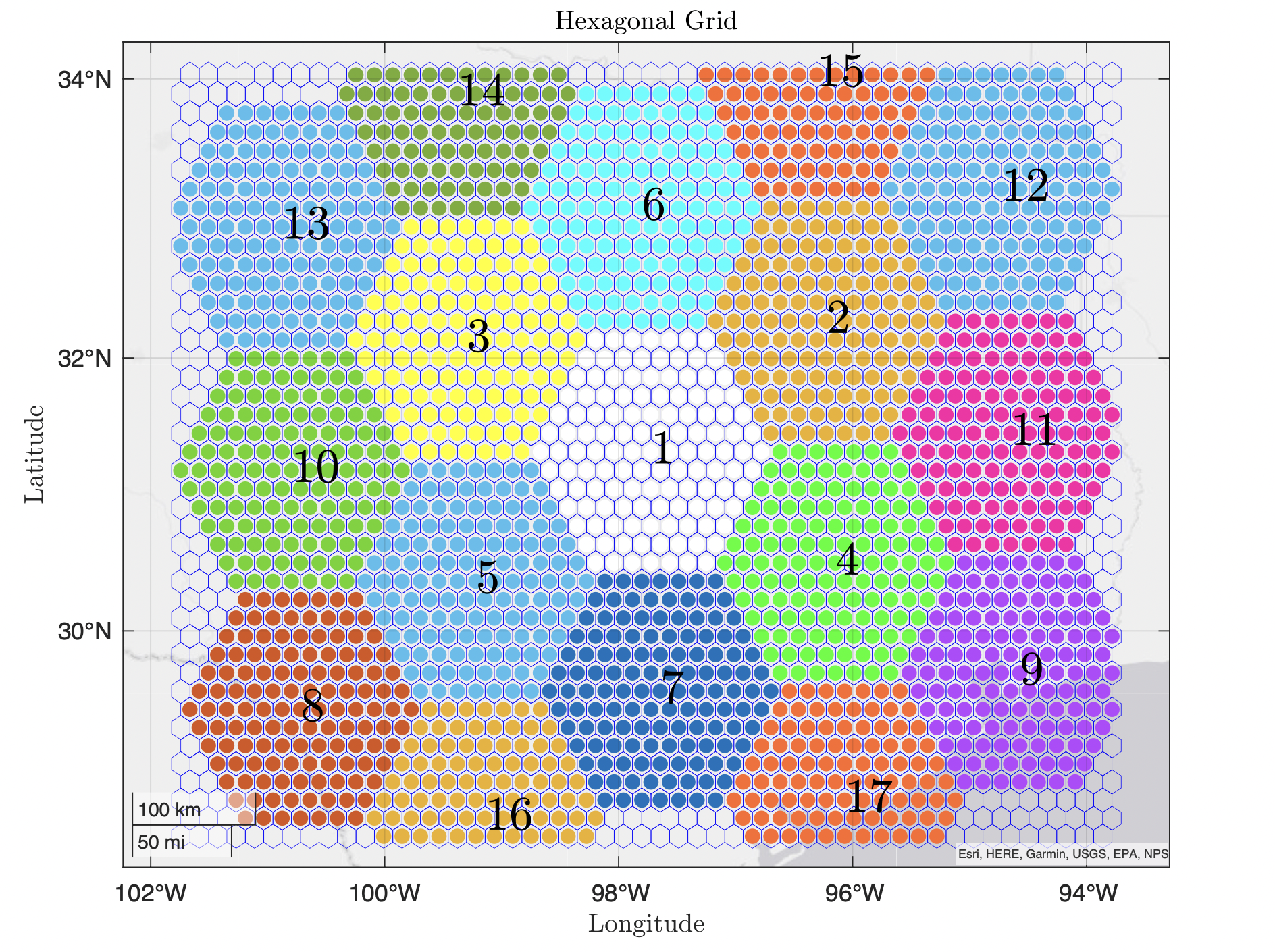} 
    \caption{An illustration of the region considered for evaluating our method. Each hexagonal element represents a cell, and each color represents a unique cell cluster. The numbers denote the priority levels assigned to each cluster.}
    \label{fig:Figure Evaluation}
\end{figure}

To evaluate our approach, we consider a scenario where the secondary system aims to predict the primary satellite assignments in the region shown in \figref{fig:Figure Evaluation}. This region consists of $N_{\mathrm{G}} = 17$ cell clusters, each with $N_{\mathrm{C}} = 127$ cells which are 100 km wide. The primary system uses the HE protocol to assign its satellites to the clusters, and the priorities of the clusters, which are unknown to the secondary system, are labeled in the figure. Furthermore, the minimum elevation angle threshold, $\epsilon_{\min}$, is set to $40^{\circ}$ \cite{spaceX_111}. In our experiments, we designate Starlink as the primary system and simulate its constellation in a Walker-Delta fashion \cite{wd} based on the orbital parameters extracted from the public filings in \cite{kuiper,spaceX_111,spacexx}. In total, we compiled a dataset of 80,647 samples to train and test our algorithm. Each sample consists of the earth-centered, earth-fixed (ECEF) and latitude, longitude, altitude (LLA) coordinates of the cells (which are the same for all samples), the ECEF and LLA coordinates of the available primary satellites, and the corresponding cell cluster associations, at a specific time instance. At any given time $t$, the number of available satellites $N_{\mathrm{P}}^t$ varies between 23 and 40, leading to the maximum availability number of $N_{\mathrm{A}} = 40$. We allocate 85\% of this dataset for training and use the remaining 15\% for testing.

We evaluate our approach against 1-layer perceptron and 3-layer (3L) multi-layer perceptron (MLP) models from \cite{coexistence_arxiv}, as well as a complete GNN characterized by fixed, fully connected directed graphs where every edge has an adjacency value of 1. The specific details of each model are provided below.

\begin{itemize}
    \item \textbf{Perceptron and 3L MLP:} Each hidden layer (if applicable) contains 2048 neurons, uses scaled exponential linear unit (SELU) activation \cite{selu}, and applies a dropout rate of 0.3. The input features include the normalized longitudes and latitudes (scaled between -1 and 1) of the available satellites and the centers of the cell clusters, along with the elevation angle between each available satellite and the center of each cluster. The models output values that are proportional to the probability of assigning a particular cluster to a specific satellite. Training is conducted using the Adam optimizer with a learning rate of 0.0001 and a batch size of 64, with the objective of minimizing the average cross-entropy loss across all clusters.
    \item \textbf{Complete GNN and GSL:} The models consist of $K = 4$ graphs and embedding layers, each outputting $F^{\prime} = 100$ features. The final representation of each node (which is the concatenation of the output features from all embedding layers) is applied a dropout rate of 0.1. In the GSL-based approach, every element of the learnable adjacency matrices is initialized using samples from a normal distribution with a mean of 0 and a variance of 0.0001. The downstream DNN classifier includes three hidden layers with 1024, 512, and 256 neurons respectively, each followed by a LeakyReLU activation with a negative slope of 0.1 and a dropout rate of 0.1. Training is conducted using the Adam optimizer with a learning rate of 0.001 and a batch size of 64.
\end{itemize}

Table I presents the top-$k$ testing accuracy of all models, calculated as the percentage of cell clusters (averaged over time) for which the correct primary satellite appears among the model's top-$k$ predictions. As shown, our GSL-based approach outperforms the baselines across top-$1$ to top-$5$ accuracy metrics. In particular, our method achieves approximately 15\% higher top-$1$ accuracy than the best baseline. A further notable observation is that, when comparing the two graph learning methods, the complete GNN performs substantially worse than the GSL-based approach. This result highlights the critical role of GSL's learnable adjacency matrices---namely, the ability to learn \emph{where} to gather information from and \emph{how much} to extract---in achieving a good classification performance.

%% file: sec-conclusion.tex
\section{Conclusion} \label{sec:conclusion}

\begin{table}[!t]
  \begin{center}
    \caption{Test Data Accuracy of Various Algorithms (\%)}
    \label{tab:table1}
    \begin{tabular}{c|c|c|c|c}
      \textbf{Accuracy} & \textbf{Perceptron} & \textbf{3L MLP} & \textbf{Complete GNN} & \textbf{GSL}\\
      \hline
      Top-$1$ & 66.58 & 61.55 & 5.95 & \textbf{80.66}\\
      Top-$2$ & 88.07 & 81.81 & 11.84 & \textbf{94.37}\\
      Top-$3$ & 95.38 & 90.17 & 17.70 & \textbf{97.76}\\
      Top-$4$ & 98.04 & 94.13 & 23.55 & \textbf{98.91}\\
      Top-$5$ & 99.05 & 96.15 & 29.35 & \textbf{99.34}\\
    \end{tabular}
  \end{center}
\end{table}

This work introduced a novel graph learning-based approach for inferring HE primary satellite assignment policies. More specifically, assuming there is access to data indicating which primary satellites have been serving which primary users over some period of time---derived from the received signal powers measured by the secondary users---we presented an end-to-end GSL-based algorithm that aims to learn the unknown priorities of the cell clusters, and, upon deployment, can directly map publicly available primary satellite coordinates into assignment decision for the clusters. Simulation results show that our method consistently outperforms baseline approaches across top-$1$ to top-$5$ prediction accuracy metrics, notably achieving around 15\% improvement over the best baseline in top-$1$ accuracy.

In future work, we aim to integrate a more explicit approach for learning the minimum elevation angle threshold $\epsilon_{\min}$, have a more direct mechanism for information sharing among the graph embedding layers, and generalize our GSL-based method beyond the HE protocol. We also intend to explore different applications where our method might be relevant.

%% file: sec-acknowledgment.tex
\section{Acknowledgment} \label{sec:acknowledgment}

This work used computational and storage services from the Hoffman2 Cluster, operated by the UCLA Office of Advanced Research Computing’s Research Technology Group.

%% file: sec-bibliography.tex
\bibliographystyle{bibtex/IEEEtran}
{\small \bibliography{bibtex/IEEEabrv,refs}}